# Trouble with the Drude Theory of Metallic Conduction: Incompatibility with Special Relativity

N R Sree Harsha[1, †], Anupama Prakash[1], A Sreedevi[2] and D P Kothari[3]

[1]*Independent Reader in Electromagnetism*

[2]*R V College of Engineering, Bangalore, India*

[3]*Gaikwad-Patil Group of Institutions, Nagpur, India*

[†]E-mail: nrsreeharshae@gmail.com

In this paper, we show that the classical Drude model of electrical conductivity, one of the fundamental models in the theory of electrical conductivity, is inconsistent with the Special Relativity. Due to this incorrect model, a current carrying closed circuit is thought not to produce second order electric field according to Maxwell's theory of electromagnetism. But, Edwards et al. detected a small second order electric field radially pointing toward a current carrying conductor in a superconducting Nb-Ti coil. Assis et al. claim to show that Maxwell's theory does not predict any second order forces and hence we should take Weber's electrodynamics seriously. But, we show that not only a magnetic field, but also a second order electric field is produced in the vicinity of a current carrying conductor, which is consistent with Maxwell's theory. This electric field points radially toward the current carrying conductor as detected in Edwards' experiments. We also show that the positive field, detected by Sansbury in a U-shaped copper conductor carrying a constant current, should be created as a consequence of our theory. We then estimate the order of the strength of this electric field and show that it is in agreement with the experimental values.

## I. INTRODUCTION

Electricity and magnetism are described completely by the Maxwell's equations of the electromagnetic field, together with the Lorentz's force-law:

$$\vec{F} = q(\vec{v} \times \vec{B} + \vec{E}) \quad (1)$$

The above equation gives the force $\vec{F}$ experienced by the charged particle with charge 'q' moving with a velocity $\vec{v}$ in the presence of an electric field described by the vector $\vec{E}$ and a magnetic field by $\vec{B}$. The electric field and magnetic field are interrelated and only depends on the frame of reference one employs to view the field. They are two aspects of same entity, which we term as the electromagnetic field [1].

However, the relation between the magnetic and the electric field was not fully understood until Einstein made his Special theory of Relativity. Only then one could see the relationship between the magnetic force on a charged particle moving near a current carrying wire and the electric force between charges [2].

It is often presumed that a current carrying conductor produces only a magnetic field in the space outside it. In other words, current carrying conductor is presumed to exert force on a charged particle near it only if there is relative motion between charge and the conductor [3]. For example, in the book *Electricity and Magnetism* by Edward M. Purcell, the linear charge densities of positive ions and electrons are considered to be equal and thus, the conductor is considered to be overall electrically neutral [4].

But, there have been many experimental demonstrations, which showed the presence of surface charges in a current carrying conductor [5], [6]. These surface charges, which are maintained by a battery, generate Electric fields inside and outside the conductor proportional to the current in the conductor. Kirchhoff first pointed this out in 1849 [7]-[9]. The three roles played by these surface charges in real circuits, as identified by Jackson [10], are: (a) they maintain the potential around the circuit, (b) they provide the electric field in the space outside of the conductor, and (c) they assure the confined flow of current by generating an electric field that is parallel to the wire. It should be emphasized that only a gradient in the surface charge density provides an electric field parallel to the wire. For an ideal conductor with zero resistance, there is no Electric field that is parallel to the wire. Hence, for an ideal conductor, there is no gradient of the surface charges on the wire [11]. This implies that the surface charges are evenly distributed along the conductor and there is only a perpendicular Electric field outside the conductor.

We have not considered radial Hall effect due to poloidal magnetic field. Due to the magnetic field inside the wire, there will be a constant negative charge density [12]. The total charge inside the wire is compensated by a positive charge density spread over the surface of the wire. This implies that radial Hall effect will not generate any electric field outside the wire. Since we are only concerned about the electric field produced outside the conductors, we do not consider this effect in our analysis.

Assis et al. calculated the force on a charged particle of charge $10^{-9}$ C due to a copper conductor carrying a current of 44.8 A [12]. They suggested that a constant current carrying conductor exerts a force on a stationary point charge placed in its vicinity. This exerted force has three components as discussed below:

- Force of attraction ($F_0$) between the point charge and the wire produced due to induction effects. A point charge will induce a distribution of charges in the nearby conductor. The net effect of these induced charges is attraction between the point charged particle and the wire. This force is independent of the current in the wire and is of order $10^{-6}$ N.

- Force ($F_1$) experienced by the point charged particle due to surface charges on the wire when a current is flowing through it. This is a first order force, which means that it is directly proportional to the drift velocity of the electrons '$v_d$' and is of order $10^{-10}$ N.

- Force ($F_2$) proportional to the square of the drift velocity of the electrons '$v_d^2$' is a second order force and is of order $10^{-15}$ N[1]. This force is dependent on the magnitude of the square of the current flowing through the conductor and is independent of its direction.

The main emphasis of this paper is the second order forces ($F_2$). Even though the second order forces are practically negligible compared to the first order forces and forces due to the electrostatic induction, they are crucial to our understanding of how currents behave in electric circuits.

The second order electric fields ($E_2=F_2/q$) produced by a current carrying conductor have not been demonstrated theoretically by Maxwell's electromagnetism. However, Weber's electrodynamics theoretically predicts the presence of the second order motional electric fields around a closed circuit carrying a constant current. Seeing the presence of the second order fields in Edwards' experiment [13], Assis et al. support the theory of electrodynamics proposed by Weber [14]-[16].

This inconsistency in the experimental evidence and the Maxwell's theory of electromagnetism is a result of an incorrect derivation of magnetic field produced by a current carrying conductor as a relativistic effect. Ron Folman first pointed out an issue in the relativistic derivation of the magnetic field in his paper [17]. However, his analysis is based on a particular impractical case when the drift velocity of the electrons is equal to the velocity of the charged particle outside the conductor. He did not quote any experiments to establish the effect.

We present an alternate derivation in section II, an estimate of the strength of the second order electric field in section III and the experimental evidence of the consequences of our derivation are presented in section IV. However, it should be emphasized that we do **not** claim this relativistic derivation of the magnetic field produced by a current carrying conductor to be new. We merely point out that this derivation is done incorrectly in the available literature and present an alternate derivation.

## II. THE THEORY OF THE SECOND ORDER ELECTRIC FIELDS

Before we discuss the alternate derivation of the magnetic field produced by a current carrying conductor, we should address few issues about the drift velocity of the electrons in a metallic conductor and the definition of current based on the idea of the drift velocity.

### 1. Definition of the current

The present definition of current in a metallic conductor is based on the Drude theory of electric conduction [18]. A simple model of an infinitely long current carrying conductor is considered in which positive ions are at rest and conduction electrons move with drift velocity "$v_d$". If the current in the conductor is "$i$", the drift velocity is given by [1]

$$v_d = \frac{i}{neA} \qquad (2)$$

Where, 'A' is the area of cross-section of the conductor, 'e' is the charge of an electron, and 'n' is number of free electrons per unit volume of the conductor. According to Ohm's law, the current '$i$' flowing through a metallic wire of resistance 'R' is proportional to the potential drop 'V' across the wire [19]. It is given by

$$i = \frac{V}{R} \qquad (3)$$

We can, immediately, see a problem with equation (2). In principle, according to equation (3), the value of the current '$i$' flowing through a conductor can be made as large as possible by decreasing the resistance of the conductor between which a constant potential difference is applied. Since the denominator in equation (2) is a constant, the drift velocity can also be made as large as possible, even greater than the speed of light. This result is inconsistent with Special theory of Relativity. Hence, we need to define the terms used in equation (2) more carefully.

We shall now present a possible solution to the above-discussed problem.

---

[1] A detailed calculation of this result is given in section III.

It is a well-known fact that if charges are in motions, the volume density of them changes because of Lorentz's contraction. The volume density changes as the time component of a 4-vector [20]. In particular the 4-vector $[\rho c, j]$, '$\rho$' being the charge density and 'j' being the current density and 'c' being the speed of light, transforms according to Lorentz's transformation. According to the available literature [1]-[4], this Lorentz's contraction in a metallic current carrying conductor is considered only when there is relative motion between charge and the conductor. However, Lorentz's contraction exists not only when there is a relative motion between the charge and the conductor but also when the current flows through the conductor. We are then lead to a new definition of current, which is illustrated as follows:

If the current in the conductor is "$i$", the drift velocity is given by equation (2). We define 'n' as the number of free electrons per unit volume of the conductor when the conductor is at rest. Since "n" changes as a time component of a 4-vector, it can also be written as

$$n = n^\dagger \gamma \qquad (4)$$

where, '$n^\dagger$' represents the number of free electrons per unit volume when the conductor is at rest and when it does not carry any current, and $\gamma = \dfrac{1}{\sqrt{1-\dfrac{v_d^2}{c^2}}}$

We can then write the equation (2) as

$$i = \lambda^\dagger v_d \qquad (5)$$

Where $\lambda^\dagger$ = neA, which is a constant for a given conductor. '$\lambda^\dagger$' also represents the electron charge per unit length of the conductor. Let, '$\lambda$' represent the charge per unit length of electrons, when there is no current through it. Then we have

$$\lambda^\dagger = \lambda \gamma \qquad (6)$$

This can be for both positive and negative charges in the conductor [21], where $\lambda_+$ is taken to be the charge per unit length of positive ions and $\lambda_-$ is taken to be the charge per unit length of electrons. Further, we assume that the charge densities are equal for a conductor carrying no current through it. Hence

$$\lambda_+ = \lambda_- = \lambda \qquad (7)$$

Thus, the point charged particle experiences no first order or second order forces when it is kept near a conductor that carries no current through it. We can write our new definition of current as

$$i = \dfrac{n^\dagger e A v_d}{\sqrt{1-\dfrac{v_d^2}{c^2}}} \qquad (8)$$

With this definition, we can see that the drift velocity of the electrons cannot be greater than 'c'. However, the value of the current can be as large as possible. We shall provide a second motivation to redefine current as show in equation (8) later in the course of development of the paper.

## 2. The physical analysis of the electric field

Let the current be "i". The electrons are moving with a drift velocity "$v_d$" in a particular direction. Consider one frame "S" which is at rest with respect to charged particle with a charge "q". The charged particle is also at rest with respect to the conductor.
In frame S,

$$\lambda'_+ = \lambda_+ \qquad (9)$$

$$\lambda'_- = \dfrac{\lambda_-}{\sqrt{1-\dfrac{v_d^2}{c^2}}} = \dfrac{\lambda}{\sqrt{1-\dfrac{v_d^2}{c^2}}} \qquad (10)$$

The primed values of $\lambda$ indicate the charge per unit length as viewed from frame "S". Thus, the charges do not balance anymore and we have a net charge developed.

This net charge developed is given by

$$\lambda' = \lambda_+ - \lambda_- = \lambda \left(1 - \dfrac{1}{\sqrt{1-\dfrac{v_d^2}{c^2}}}\right) \qquad (11)$$

$$\lambda' = \lambda(1-\gamma) \qquad (12)$$

Where, $\gamma = \dfrac{1}{\sqrt{1-\dfrac{v_d^2}{c^2}}}$

Thus, as $\gamma > 1$, a negative electric field is produced radially because of this inequality of the positive and negative charge distributions. The strength of this field, at a distance 'r' from the conductor, is given by [2]:

$$E = \dfrac{\lambda'}{2\pi\varepsilon r} = \dfrac{\lambda(1-\gamma)}{2\pi\varepsilon r} \qquad (13)$$

where, "$\varepsilon$" represents the permittivity of space.

Accordingly, the force on the charged particle with charge "q" due to this field is [22]

$$F = qE \qquad (14)$$

$$F = \dfrac{\lambda(1-\gamma)q}{2\pi\varepsilon r} \qquad (15)$$

To show that the force given by equation (15) is indeed a second order force in drift velocity '$v_d$' to the first approximation, we write it as

$$F \approx \frac{\lambda q}{2\pi\varepsilon r}\left[1-\left(1+\frac{v_d^{\,2}}{2c^2}\right)\right] \quad (16)$$

using the approximate binomial expansion and the fact that practically, $v_d \ll c$. In fact, it is now known that $v^2/c^2 < 10^{-20}$ for essentially all cases for metallic conductors at room temperatures [13]. Thus, the equation (16) is a valid approximation of the equation (15).
Equation (16) can then be reduced to

$$F \approx -\frac{\lambda q v_d^{\,2}}{4\pi\varepsilon r c^2} \quad (17)$$

Equation (17) shows that the current carrying conductor produces a small amount of electric field pointing radially toward the wire that is of second order to the first approximation. In the next section, we present a complete derivation of the magnetic field produced in a current carrying conductor as a relativistic effect.

### 3. Magnetism as a relativistic effect

Let the velocity of the charged particle be "u" with respect to the conductor. We assume here, for simplicity, that the charged particle moves to the left (negative direction) and the current "i" has the opposite direction, i.e. moves to the right (positive direction). Refer FIG. 1 and 2. Consider a frame "S" which is at rest with respect to the charged particle and frame "S†" which is at rest with respect to the conductor.

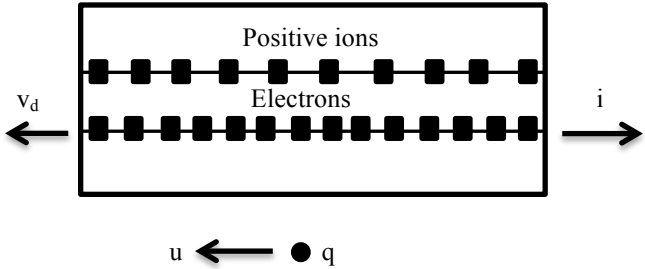

FIG. 1. Conductor as seen from its own rest frame "S†".

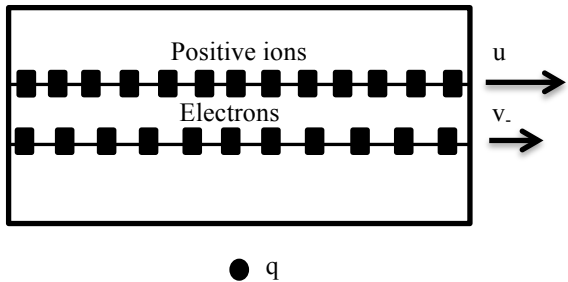

FIG. 2. Conductor as seen from the frame "S".

From the frame "S", the positive ions move with the velocity "u" in positive direction and also the conduction electrons move with velocity "v_-", in the same direction, given by

$$v_- = \left(\frac{u - v_d}{1 - \frac{u v_d}{c^2}}\right) \quad (18)$$

At this point, we define $\gamma_1 = \dfrac{1}{\sqrt{1-\dfrac{u^2}{c^2}}}$.

As positive ions move with the velocity "u" with respect to charge, the charge distribution of it changes and is given by

$$\lambda_+ = \left(\frac{\lambda}{\sqrt{1-\dfrac{u^2}{c^2}}}\right) = \lambda \gamma_1 \quad (19)$$

And similarly,

$$\lambda_- = \frac{\lambda}{\sqrt{1-\dfrac{v_-^{\,2}}{c^2}}} \quad (20)$$

Substituting $v_-$ from the equation (18) in equation (20), we get

$$\lambda_- = \lambda \gamma \gamma_1 \left(1 - \frac{u v_d}{c^2}\right) \quad (21)$$

Thus, the total charge density as seen from frame "S" is:

$$\lambda' = \lambda_+ - \lambda_- = \lambda \gamma_1 \left(1 - \gamma + \frac{u v_d \gamma}{c^2}\right) \quad (22)$$

The force on the charged particle with charge "q" placed at a distance "r" from the conductor is given by

$$F = \gamma_1 \left(\frac{\lambda\left(1-\gamma+\dfrac{u v_d \gamma}{c^2}\right)}{2\pi\varepsilon r}\right) q \quad (23)$$

But, we need force "$F_0$" in the rest frame of the conductor (frame "S†") so as to relate it to good old familiar form. The force "F" in frame "S" is to be converted to frame "S†" by

$$F_0 = \frac{F}{\gamma_1} \quad (24)$$

Thus, from equation (23), the force on the particle is

$$F_0 = \left( \frac{\lambda\left(1 - \gamma + \frac{uv_d\gamma}{c^2}\right)}{2\pi\varepsilon r} \right) q \quad (25)$$

With the permeability of space denoted by "μ", the speed of light is given according to Maxwell as

$$c = \frac{1}{\sqrt{\mu\varepsilon}} \quad (26)$$

Substituting the above formula in equation (22) we get,

$$F_0 = Buq + \frac{q\lambda(1-\gamma)}{2\pi\varepsilon_0 r} \quad (27)$$

where, $B = \frac{\mu i}{2\pi r}$ is the known formula used to calculate the strength of the magnetic field at a distance "r" from the conductor. Note that we have used the new definition of current defined in section II in the equation (27). The second motivation to do this is the fact that the Lorentz's force form, as given by equation (1), is relativistically correct and to preserve its form, we should redefine current as shown in section II.

The equation (27) shows that when a current flows through a conductor, not only magnetic field but also a second order (to first approximation) electric field is created. The first term and the second term represent the magnetic field and the electric field produced respectively. The equation (27) can be represented in vector form as follows

$$\vec{F} = q\left[\left(\vec{u} \times \vec{B}\right) + \vec{E}(1-\gamma)\right] \quad (28)$$

Where, $E = \frac{\lambda}{2\pi\varepsilon r}$

We can also write equation (27) in Lorentz's form as

$$\vec{F} = q\left[\left(\vec{u} \times \vec{B}\right) + \vec{E}^\dagger\right] \quad (29)$$

Where, $\vec{E}^\dagger = \vec{E}(1-\gamma)$

Thus, the electric field is created because of charge imbalance between positive ions and electrons caused due to the flow of the current along with the magnetic field.

## III. AN ESTIMATE OF THE STRENGTH OF THE ELECTRIC FIELD

We present a simple estimate of the electric field in the vicinity of the copper wire, which carries a current of 44.8 A. The approximate density of the conduction electrons in copper, when there is no current flowing through it, is $n^\dagger = 8.45 \times 10^{28} / m^3$.

For a copper with a cross-section area A = 1 mm$^2$, carrying a current i = 44.8 A, with the charge of a single electron e = 1.6×10$^{-19}$C, the electronic drift velocity is given by

$$v_d = \frac{i}{neA}$$

$$\frac{v_d}{\sqrt{1 - \left(\frac{v_d}{c}\right)^2}} = \frac{i}{n^\dagger eA}$$

$$v_d \approx 3.31 \times 10^{-3} m/s$$

With this drift velocity, we get

$$\gamma \approx \left[1 + \frac{v_d^2}{2c^2}\right] \approx 1.6 \times 10^{-22}$$

The iconic or electronic charge $\lambda$ is given by

$$\lambda = n^\dagger eA = (1.6 \times 10^{-19})(8.45 \times 10^{28})(10^{-6})$$

$$\lambda \approx 1.4 \times 10^4 C/m$$

The electric field at a distance of 1 cm from this wire is therefore given by

$$E \approx -\frac{\lambda v_d^2}{4\pi\varepsilon_0 rc^2} \approx 1.53 \times 10^{-6} N/C$$

If the charge of the particle is "10$^{-9}$ C", the magnitude of the force due to this electric field is

$$F_E \approx 1.5 \times 10^{-15} N$$

For comparison, the gravitational force on a particle of mass 1 microgram is $10^{-8} N$.

Thus, the force experienced by the charged particle due to the second order force produced by a current carrying conductor is so minute as not to be easily detectable. This is because of the fact that, in practice, the drift velocity of the electrons is very low and hence $\gamma$ is nearly equal to 1.

## IV. EXPERIMENTAL EVIDENCE

The experiments conducted by Edwards et al. and Sansbury suggest an additional component of the electric field outside a current carrying conductor, which cannot be explained by existing theory and thus is a subject for investigation. However, the proposed second order field seems to provide some justification.

### 1. Edwards' experiment

In the paper demonstrating the experimental evidence for second-order force from a stationary current carrying conductor, Edwards' et al. show two derivations from Maxwell's theory that show that E=0 [13]. In this section, we present a brief review of one of the derivations and show how the second-order fields produced due to Lorentz's contraction is consistent with Maxwell's theory.

The electric field, $\vec{E}$, resulting from charges and currents is given by

$$\vec{E} = -\nabla\phi - \frac{\partial \vec{A}}{\partial t} \qquad (30)$$

where, $\phi$ and $\vec{A}$ are the usual retarded potentials.

For steady currents, the distribution of currents along the wire is uniform. Thus, in a charge neutral circuit, $\phi = 0$. As Baker has shown in a stationary circuit if the currents are constant, then $\frac{\partial \vec{A}}{\partial t} = 0$. Hence, $\vec{E} = 0$.

As we can see, the second order electric fields we have derived are due to charge imbalance in the conductor resulting from Lorentz's contraction. Thus, linear charge densities of electrons and positive ions are not equal in the circuit carrying constant current. This ensures that Gauss's law is not violated. This imbalance in charge densities can also explain the variation II in the experiment demonstrated by Edwards et al., A super-conducting coil was used and was galvanically isolated from the electrometer. The diagram is as shown in FIG. 3.

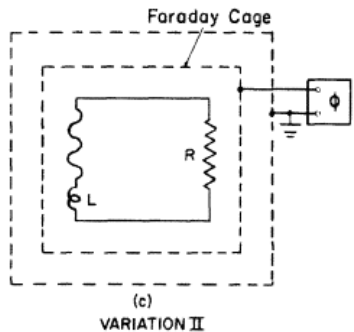

FIG. 3. Simplified circuit diagram of an experiment conducted by Edwards et al. (shown as FIG. 4. in their paper).

The team has reported an $I^2$ signal even in this condition. They have concluded that the field has dependence similar to that of electromagnetic fields. We now know that the charge imbalance in the current carrying conductor produced an electric field consistent with our derivation.

We have only considered an infinitely long straight current carrying ideal conductor in all our analysis. We shall now attempt to generalize it to include circuits which not straight. Consider a perfect circular closed metallic conductor with a constant current flowing through it. The positive ions are at rest and the conduction electrons are moving with a constant speed but are continually experiencing force toward the center of the circular geometry by the distribution of surface charges maintained by the battery. We then ask the question: is the space between the electrons Lorentz's contracted? The situation is similar to rotating rigid disk. The radius of the rotating disk always moves perpendicular to the circular motion of the elements at the periphery hence its length should remain the same as seen from an inertial frame. However, the periphery is Lorentz's contracted. This is called Ehrenfest's paradox [23] and played a key role in establishing the idea that the geometry in non-inertial frames of references is non-Euclidean. Later, Einstein resolved the paradox showing that the geometry is non-Euclidean [24]. We shall not go further into these details except to state that the disk does not appear to Lorentz's contract as viewed from an inertial rest frame of the disk, which is beautifully given in Øvind Grøn's book [25]. Similarly, the space between the electrons does not appear Lorentz's contracted as viewed from an inertial rest frame of the conductor.

However, if the conductor is not a perfect circular but like a solenoid, the electric field produced outside the coil due to Lorentz's contraction, can be taken as superposition of fields generated due to a stack of circular rings and a longitudinal current parallel to the length of the coil [see figure 6.20 of ref. 4]. Since, stack of circular rings do not produce any electric fields of second order, we can say that the electric field produced by a solenoid is completely equivalent to the electric field produced by a straight wire of length equal to length of the solenoid. This, maybe, was the field detected in Edwards' experiment.

### 2. Sansbury's experiment

We now consider a U-shaped circuit as shown in FIG. 4.

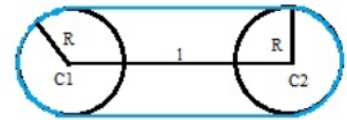

FIG. 4. A circuit to demonstrate results obtained by Sansbury.

The conductor (shown in blue) is a closed circuit and is wound on two circular wheels whose centers "C1" and "C2" are 'l' meters apart. The radii of both the circular wheels are equal to 'R'. The total length of the conductor then is equal to $2(l+\pi R)$. We shall analyze the circuit in the inertial rest frame of the conductor. As shown in previous section, Lorentz's contraction only occurs in the parts where the conductor is straight. Let 'λ' denote the magnitude of the linear charge densities of both positive ions and electrons when there is no current in the conductor. But, when a current starts flowing through the conductor, the linear charge densities of electrons in the linear portions of the wire increases as discussed in section III. Thus, due to charge imbalance, a small second order electric field is generated in the linear portions of the wire. However, the circular portions of the wire do not introduce any change in charge densities. Let the linear charge density of the electrons be $\lambda'_l$ in the linear portion of the conductor and $\lambda'_c$ in the circular portion of the conductor when the current is flowing. For the conservation of charge, we require that

$$\lambda(2l + 2\pi R) = \lambda'_l(2l) + \lambda'_c(2\pi R) \qquad (31)$$

This immediately gives us the value of $\lambda'_c$ as

$$\lambda'_c = \lambda \frac{(l - l\gamma + \pi R)}{\pi R} \qquad (32)$$

The total charge in a single circular portion '$Q_c$' of the wire is then given by

$$Q_c = \lambda \pi R - \lambda'_c \pi R \qquad (33)$$

Substituting for $\lambda'_c$ from equation (32) in the equation (33), we get

$$Q_c = \lambda l (\gamma - 1) \qquad (34)$$

Thus, the circular portions of the wire become positively charged and create an electric field, which is of second order to the first approximation. In an experimental demonstration, Sansbury found a force between stationary charge on a metal foil and a steady electric current in a U-shaped copper conductor [26]. He wrote

*"It appears the perfect electrostatic screening of metal ions in a copper conductor by the conduction electrons is **somehow** upset by the flow of an electric current. The screening deficiency makes the conductor appear to possess a net positive charge."*

The copper conductor was positively charged when the current was flowing through it as expected by equation (34). The positive charge in the U shaped conductor was found to be independent of the direction of the current. This eliminates the possibility that the forces detected by Sansbury in his experiment are of first order. If the forces were of first order, the conductor should be negatively charged when the current is made to flow in the reverse direction. However, this was not observed to be the case. A more detailed analysis is required to confirm that the positive charge is a second order effect, to the first approximation. This can be a good line of experimental research.

## V. DISCUSSIONS AND CONCLUSION

It has been shown that a straight current carrying conductor produces a small electric field of second order to the first approximation. The strength of the electric field is very small and further sensitive experiments are needed to calculate the effects produced by the field. The effects presented in this paper are observable only when large currents, consecutively large drift velocity of the electrons, are involved. It is not required to invoke Weber's electrodynamics to explain the second order fields produced by the conductor carrying current. A small amount of negative electric field should be produced as a consequence of Special Relativity and charge invariance.